# Metal – insulator transition in $CaV_{1-x}W_xO_3$ (x = 0.1-0.33) perovskites

I.V. Morozov,[1] I.K. Shamova,[2] M.A. Yusifov,[1,3] S.Y. Istomin,[1,4] T.B. Shatalova,[1] A.I. Boltalin,[1] A.A. Andreev,[5] R.G. Chumakov,[5] T.M. Vasilchikova,[2,6] A.A. Fedorova,[1] E. A. Ovchenkov,[2,6] O.S. Volkova[2,6]

[1]Department of Chemistry, M. V. Lomonosov Moscow State University, 119991 Moscow, Russian Federation
[2]Functional Quantum Materials Laboratory, National University of Science and Technology "MISiS", Moscow 119049, Russia
[3]Department of Chemistry, Lomonosov Moscow State University, Baku Branch, AZ1146 Baku, Azerbaijan
[4]HSE University, Moscow, Russia
[5]National Research Centre "Kurchatov Institute". 123182, 1 Kurchatov Square, Moscow
[6]Department of Physics, M. V. Lomonosov Moscow State University, 119991 Moscow, Russian Federation

Novel $CaV_{1-x}W_xO_3$ ($0.1 \leq x \leq 0.33$) oxides with an orthorhombically distorted perovskite structure of $GdFeO_3$ type have been synthesized. These compounds contain in B-position $W^{+6}$ and V cations in an oxidation state between +4 ($CaVO_3$) and +3 ($x$=0.33). $CaV_{0.9}W_{0.1}O_3$ compound possesses metallic type of conductivity and Pauli paramagnetism. The intermediate compositions are between bad metal and semiconducting type of behavior with paramagnetic response. $CaV_{0.67}W_{0.33}O_3$ is a Mott insulator with localized $V^{+3}$ moments coupled by strong antiferromagnetic interactions. It demonstrates the reduction of effective magnetic moment at high temperatures and canonical spin glass state formation with the freezing temperature $T_g$ = 27.5 K seen in dc - and ac - magnetic susceptibility. Disorder in the magnetic subsystem induces a broad peak in magnetic contribution of the heat capacity at $T_{max}$ = 46 K.

**Keywords** metal – insulator transition, Mott insulator, perovskite, magnetic $V^{+3}$

Mott insulators are demanded for the development of neural networks and non-volatile memory elements [1,2]. Their transport properties are discussed within the model of Zaanen, Sawatzky and Allen where the *d-d* Coulomb repulsion energy $U$ is compared with charge-transfer energy $\Delta$ between *p*-level of the ligand and *d*-level of the metal. In materials with $U < \Delta$, the band gap equals to $U$ and they are Mott – Hubbard insulators. If $\Delta < U$, the band gap is given by $\Delta$ and materials are charge transfer insulators or metals [3].

The $ABO_3$ perovskites are a good platform to search for new Mott insulators due to superior flexibility for chemical substitutions including almost all elements of the Periodic Table and several existing examples of its implementation [4,5]. In this family $CaVO_{3-\delta}$ stands for the closeness to the metal insulator transition [6]. It formally contains $V^{+4}(d^1)$ ions in an octahedral oxygen coordination and exhibits Pauli paramagnetism [7]. It demonstrates both insulating and metallic type of conductivity depending on stoichiometry of oxygen composition [7,8]. Photoemission spectroscopy shows the formation of Hubbard bands in $CaVO_{3-\delta}$ [9]. Its XAS spectra indicate a charge transfer regime, i.e. $\Delta < U$ [10]. Heterovalent substitution in A position of $Ca^{+2}$ for rare earth ions $RE^{+3}$ (RE = La, Y) leads to the metal – insulator transition at the concentration of lanthanum and yttrium equal to 0.8 and 0.5 [11-13]. Pure $LaVO_3$ and $YVO_3$ are antiferromagnetic Mott-insulators with two electrons in the *d*-band [14,15]. DFT calculations revealed importance of steric factor for the formation of a localized state in $LaVO_3$. It lowers the energy in the high temperature paramagnetic state admitting rotations of oxygen octahedra and antipolar motions, which results in trapping of charge carriers [16].

Isovalent replacement of V by $Ti^{+4}$ produces strong correlation fluctuations giving a smooth metal-insulator transition for titanium content between 20-40% [17]. While $CaV_{1-y}Mo_yO_3$ ($y$ = 0.2-0.6) system exhibits only metallic type of behavior with a Pauli – type paramagnetic response, most likely due to the formation of a mixed energy band between Mo and V [18]. For $CaV_{1-x}W_xO_3$, the situation is expected to be different, since the W 5d orbitals are higher in energy than the V 3d orbitals. Therefore, no mixed band is expected for these perovskites. Theoretically it was predicted that if a double perovskite $CaV_{0.5}W_{0.5}O_3$ with ordered V and W cations in the crystal structure could be prepared, it might demonstrate antiferromagnetism in an insulating regime [19]. In present work we report on the synthesis of $CaV_{1-x}W_xO_3$, $x$=0.1-0.33 compounds. We have provided their structural characterization and studied basic physical properties, i.e. specific heat, magnetization, and resistivity and found Mott insulator behavior for the extreme composition with $x$ = 0.33.

**Materials and methods**

A modified citrate method was used to prepare $CaV_{1-x}W_xO_3$ ($x$ = 0.1 - 0.33) samples. At the first stage, starting substances, i.e. $Ca(NO_3)_2{\cdot}4H_2O$, $V_2O_5$, $(NH_4)_{10}W_{12}O_{42}{\cdot}5H_2O$ were dissolved in an aqueous solution of citric acid, with the addition of ammonium nitrate. The obtained solution was evaporated at 130-140°C followed by heating up to 200-250°C until it transformed into a solid. At the second stage, the obtained solid product was calcinated in air at 600°C for 12 hours, thoroughly ground in a mortar and pressed into a pellet. At the third stage, yellow-white pellets were annealed in a reducing atmosphere of 5%$H_2$/Ar at a temperature of 1350-1400°C for 10 hours. The prepared samples were kept in a dry box to prevent the contact with air moisture.

Phase composition of the samples was determined by X-ray powder diffraction (XRPD) with a Huber G670 Image plate Guinier diffractometer ($CuK_{\alpha1}$ radiation, curved Ge monochromator) and STOE STADI-P diffractometer with $CuK_{\alpha1}$ radiation using a Si (111) curved monochromator. Refinement of the crystal structure of $CaV_{0.67}W_{0.33}O_3$ was performed using the GSAS program package [20]. X-ray photoelectron spectra (XPS) of $CaV_{0.67}W_{0.33}O_3$ were obtained using a hemispherical electron analyzer PHOIBOS 150 with excitation by monochromatized Al Kα radiation (photon energy 1486.61 eV and resolution ΔE = 0.2 eV), which is located at the NanoPES beamline of Kurchatov Synchrotron-Neutron Research Complex (National Research Center "Kurchatov Institute") [21]. Powdered samples were pressed into carbon tape on a sample holder and then transferred to the spectrometer vacuum chamber with a base pressure of $3{\cdot}10^{-9}$ mbar. All spectra were measured in constant transmission energy mode with energies of 120 and 60 eV for the overview spectra and fine structure of individual lines, respectively. To analyze the experimental data and decompose the lines into components, the CasaXPS software [22] was used, as well as Ref. [23,24] to interpret the data obtained. The oxygen content in $CaV_{1-x}W_xO_3$ samples was determined by thermogravimetry using NETZSCH STA 449 F3 Jupiter analyzer. The samples were heated in an air atmosphere at a rate of 10ºC/min up to 1050 °C and holded at this temperature for 1 hour. The composition of the gas phase was controlled using a quadrupole mass spectrometer QMS 403 Quadro (Netzsch, Selb, Germany). X-ray microanalysis was performed by means JEOL JSM 6490 LV scanning electron microscope operating at 30 kV, equipped with an energy-dispersive X-ray analysis system INCA Energy+, Oxford Instruments (UK).

Temperature dependences of resistivity of $CaV_{1-x}W_xO_3$ were measured using the four-point method on polycrystalline sintered samples using Resistivity option of Physical Property measurement system "Quantum Design" (PPMS 9T). Magnetic properties were measured by Magnetic Property Measurement System 7T and ACMS option of PPMS 9T. Thermal properties of $CaV_{0.67}W_{0.33}O_3$ were measured by Heat Capacity option of PPMS 9T.

**Results and discussion**

**Crystal structure**

Black samples of $CaV_{1-x}W_xO_3$ ($x$ = 0.1 - 0.5) were successfully prepared after final annealing in a reducing atmosphere. According to scanning electron microscopy, the obtained samples are well sintered and consist of faceted grains with linear dimensions of 1–6 μm. The powder X-Ray diffraction patterns of $CaV_{1-x}W_xO_3$ are shown in left panel of Fig. 1. The splitting of the perovskite subcell reflections together with the presence of superstructure reflections indicates for the formation of an orthorhombically distorted perovskite of $GdFeO_3$ type. A gradual shift of reflections towards smaller angles with the increase of tungsten content indicates for the formation of solid solutions. Estimated unit cell parameters and the volume of the unit cell of $CaV_{1-x}W_xO_3$ compounds are presented in Table. S1 and right panel of Fig. 1. They increase for higher $x$ reaching a maximum value for $x = 0.33$. For $x>0.33$ the unit cell volume does not change. Thus, the limiting level of substitution is $x = 0.33$. This observation correlates with additional impurity $CaWO_4$ peaks seen in left panel of Fig. 1 for $CaV_{0.5}W_{0.5}O_3$.

To evaluate qualitatively the oxidation state of vanadium and tungsten in the limiting composition $CaV_{0.67}W_{0.33}O_3$, the XPS spectra, shown in Fig. S1 and S2, were obtained. As can be seen from Fig. S1, the spectrum of vanadium contains three peaks, close in binding energies to $V^{+2}$ at 513.5 eV, $V^{+3}$ at 515.5 eV and $V^{+4}$ at 517.1 eV [24]. Thus, average oxidation state of vanadium is close to +3. Fig. S2 shows the spectrum of W 4d. The spectrum of W 4d is presented as a double doublet with a binding energy of 244.8 eV and 247.6 eV for the W $4d_{5/2}$ lines. According to Refs. [25,26], these binding energies of the W $4d_{5/2}$ lines correspond to $W^{+5}$ and $W^{+6}$. It is worth noting that these concentrations differ from the bulk value. Partial reduction of tungsten is mostly likely due to the high vacuum conditions during the XPS experiment. Qualitatively XPS method indicates for the presence of $V^{+3}$ and $W^{+6}$ in $CaV_{0.67}W_{0.33}O_3$.

Formally, replacement of each $V^{+4}$ by $W^{+6}$ in $CaVO_3$ leads to the reduction of two $V^{+4}$ cations to $V^{+3}$ according to the chemical formulae $Ca\{[V^{+4}]_{1-3x}[V^{+3}]_{2x}[W^{+6}]_x\}O_3$. From this formula, one can estimate the extreme composition of a solid solution as $x$=0.33. $CaV_{0.67}W_{0.33}O_3$ contains only $V^{+3}$ in B position. The cation composition of the samples was confirmed by EDX analysis data and anion composition was derived from TGA data as shown in Fig. S3 and Table S1. Appearance of large $V^{+3}$ and $W^{+6}$cations with r($V^{+3}$)=0.78 Å and r($W^{+6}$)=0.74 Å instead of small $V^{+4}$ with r($V^{+4}$)=0.72 Å resulted in the increase of the unit cell volume (Table S1, Fig. S1) and average B-cation size. The latter lead to a decrease in the tolerance factor ($t$-factor) and an increase in the orthorhomic distortion of the crystal structure, expressed as $(a - c)/(a + c)$, which changes from 0.0011 for $x = 0.1$ to 0.0063 for $x = 0.33$.

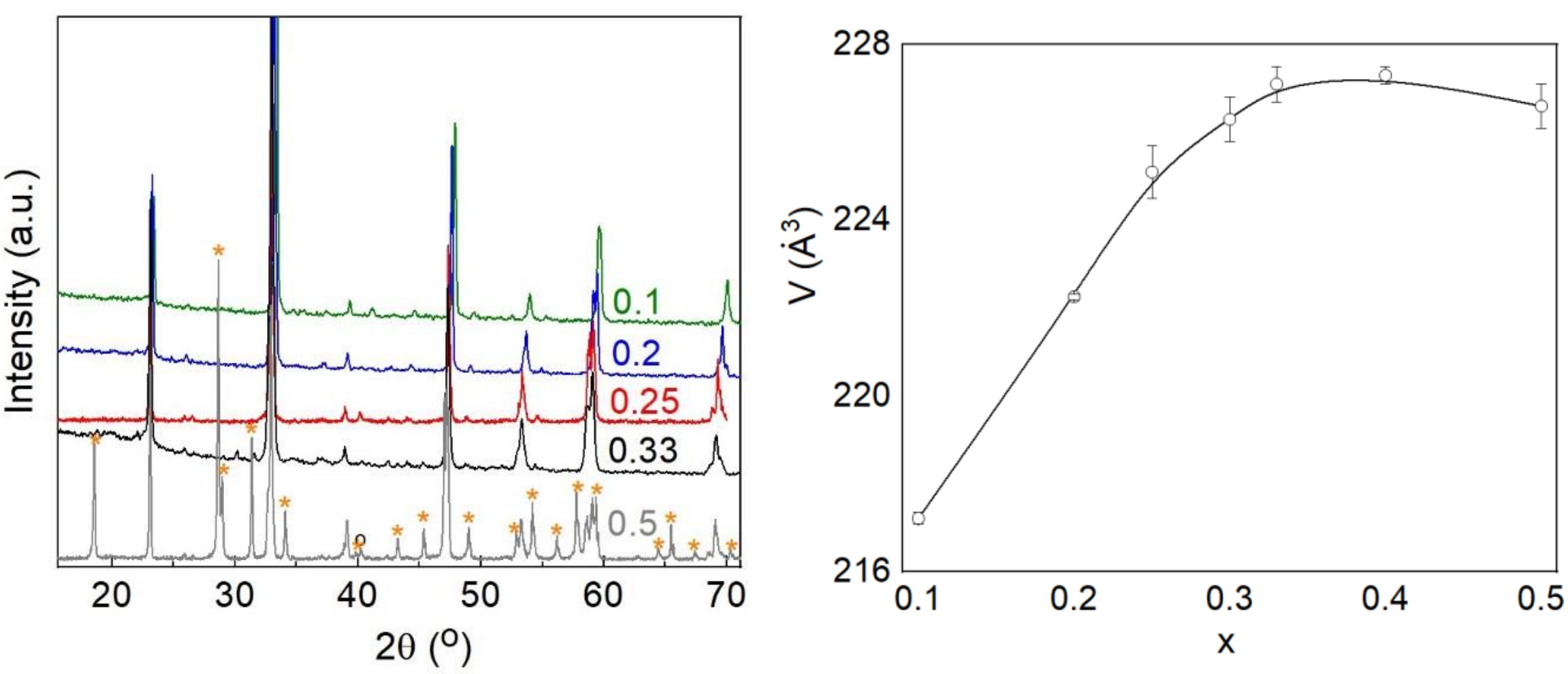

Fig. 1 Left: Powder X-Ray diffraction patterns of $CaV_{1-x}W_xO_3$ samples with $x$ = 0, 0.1, 0.2, 0.25, 0.33 and 0.5. Stars and open circle mark impurity peaks of $CaWO_4$ (ICDD PDF #72-257) and metallic W (ICDD PDF #4-806). Right: Dependence of the unit cell volume of $CaV_{1-x}W_xO_3$ perovskite on $x$.

The crystal structure of $CaV_{0.67}W_{0.33}O_3$ was refined using powder X-Ray diffraction data in the model assuming the absence of B-cation ordering (*i.e.*, in the $GdFeO_3$ type structure (space group *Pnma*) (Table 1, Fig. 2)). The occupancy of the B-position by V and W was fixed according to the nominal composition and was not refined. The atomic displacement parameters for oxygen atoms were refined in the block. Refined mass fraction of the perovskite phase is 98.6(1)%. The sample contains a relatively small amount of impurity phases, namely 1.1(3) wt. % of $Ca_3WO_6$ and 0.2(1) wt. % of metallic W. However, these impurities are nonmagnetic and their presence does not affect the magnetic behavior of the main phase. Refinement in the model assuming the ordering of V and W in the crystal structure (space group *P2₁/n*) did not prove the refinement. The absence of B-cation ordering in $CaV_{0.67}W_{0.33}O_3$ is most likely due to rather close ionic radii of $V^{+3}$ and $W^{+6}$ [27].

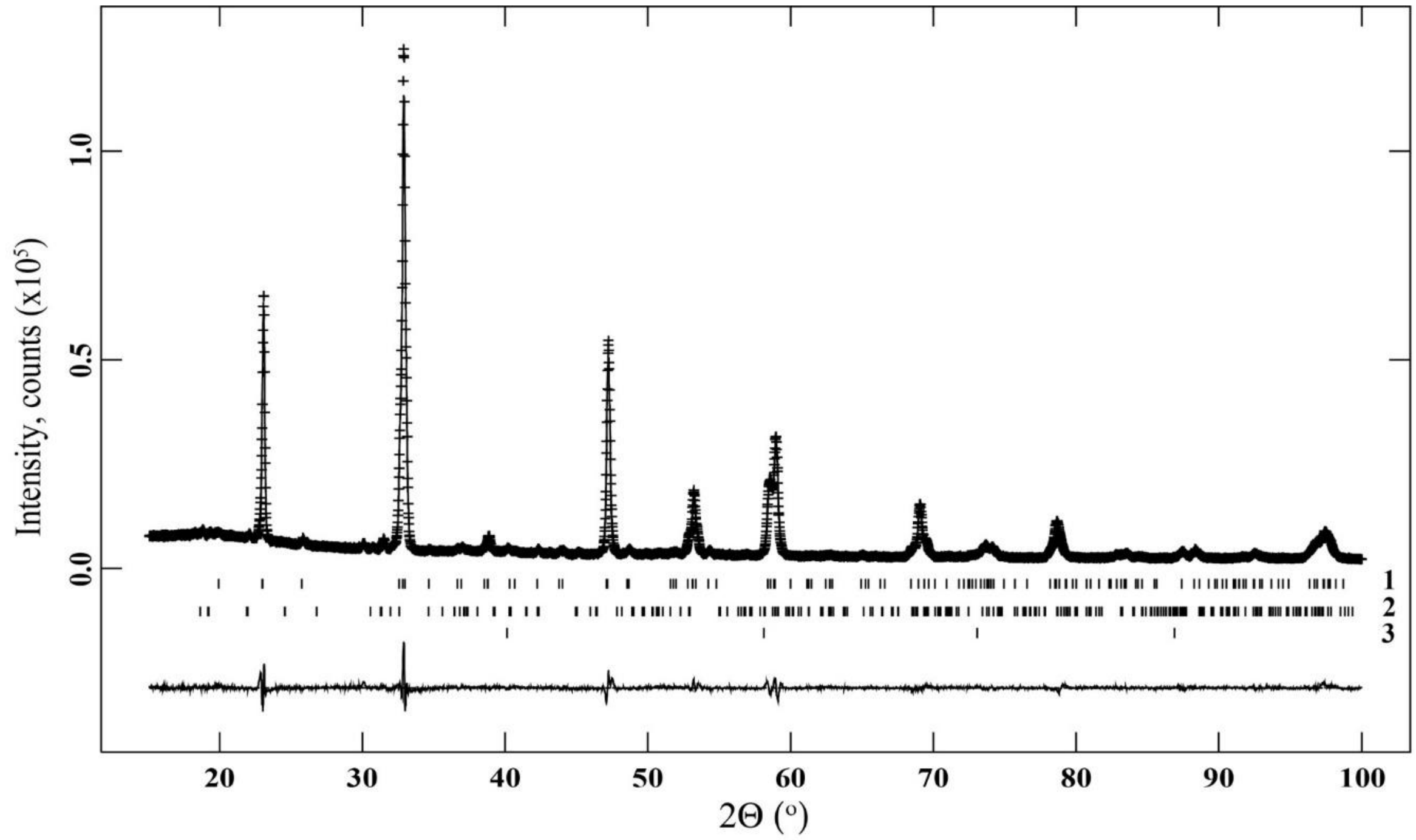

Fig. 2. Observed, calculated and difference X-ray powder diffraction profiles for $CaV_{0.67}W_{0.33}O_3$ sample along with reflection positions of $CaV_{0.67}W_{0.33}O_3$ (1), $Ca_3WO_6$ (2) and W (3) phases.

Table 1 Atomic coordinates and displacement parameters in the crystal structure of $CaV_{0.67}W_{0.33}O_3$ (space group *Pnma,* *a*=5.47104(9), *b*=7.6806(1), *c*=5.40602(8) Å; $R_p$=0.0420, $R_{wp}$=0.0569, $\chi^2$=17.0).

| Atom | x | y | z | $U_{iso}$, Å$^2$ |
|---|---|---|---|---|
| Ca | -0.0365(4) | 0.25 | -0.0064(9) | 0.0275(4) |
| V/W* | 0 | 0 | 0.5 | 0.0229(2) |
| O1 | 0.521(1) | 1/4 | 0.075(1) | 0.023(1) |
| O2 | 0.2068(8) | 0.0449(7) | -0.218(1) | 0.023(1) |

* Position occupancy is 0.67 for V and 0.33 for W

The coordination polyhedron of the B-cation (B=V, W) in the crystal structure of $CaV_{0.67}W_{0.33}O_3$ is *i*-centered [$BO_6$] octahedron distorted as shown in Table 2: interatomic B-O distances in the [$BO_6$] octahedron vary in the range 1.931(5)-2.019(5) Å. The bond angle between the corner-shared (V/W)$O_6$ octahedra amounts (153.6(2)-155.3(4)°), which is significantly lower than in the cubic perovskite structure. Average B-O bond length in $CaV_{0.67}W_{0.33}O_3$ is 1.97(4) Å, which is larger than 1.914(5) Å of $V^{+4}$-O bond in $AEVO_3$ (AE – alkali earth element) but shorter than 2.005(8) Å of $V^{+3}$-O bond in $LnVO_3$ (Ln – rare earth element) perovskites. The B-O bond length is close to the value estimated as $2/3\times l(V^{+3}\text{-O}) + 1/3\times l(W^{+6}\text{-O}) = 1.98$ Å where metal – oxygen bonds are taken from Inorganic Crystal Structure Database [28].

Table 2. Selected interatomic distances (Å) and angles (°) in the crystal structure of the $CaV_{0.67}W_{0.33}O_3$ perovskite.

| | | | | | |
|---|---|---|---|---|---|
| Ca-O1 | 2.463(7) | | B-O1 | 1.966(1) x2 | |
| O1 | 2.353(8) | | O2 | 1.931(5) x2 | |
| O2 | 2.358(5) x2 | | O2 | 2.019(5) x2 | |
| O2 | 2.733(6) x2 | | | | |
| O2 | 2.584(6) x2 | | | | |
| | | | | | |
| B-O1-B | 155.3(4) | | | | |
| B-O2-B | 153.6(2) | | | B=$V_{2/3}W_{1/3}$ | |

**Resistivity, magnetic properties and specific heat**

Temperature dependences of resistivity ρ of $CaV_{1-x}W_xO_3$ are shown in Fig. 3. Metallic type of resistivity for $x$=0.1 is replaced to the insulating one for $x$ = 0.33 as is shown in left panel of Fig. 3. The ρ(T) of $CaV_{0.9}W_{0.1}O_3$ is described with $\rho(T) = \rho_0 + AT^2$ function in a broad temperature range with $\rho_0 = 8.6 \cdot 10^{-5}$ Ω·cm, $A = 6.2 \cdot 10^{-10}$ Ω·cm/K$^2$ as is shown in Fig. S4 in Supplementary section. Obtained values of $\rho_0$, $A$ and room temperature resistivity $1.3 \cdot 10^{-4}$Ω·cm are comparable with that in crystals of $CaVO_3$ [29]. The $CaV_{0.80}W_{0.20}O_3$ and $CaV_{0.75}W_{0.25}O_3$ are between bad metal and semiconducting type of behavior. Their room temperature values of resistivity are $1.4 \cdot 10^{-3}$ and $7.6 \cdot 10^{-3}$ Ω·cm. The $CaV_{0.67}W_{0.33}O_3$ has four orders of magnitude higher resistance at room temperature equal to 1.4·10 Ω·cm and shows a rapid increase in resistivity with lowering temperature. The dependence of room temperature resistivity vs. $x$ is shown in the inset to Fig. 3. It demonstrates a deflection at $x_C = 0.2$.

The resistivity of $CaV_{0.67}W_{0.33}O_3$ demonstrates two activation regimes seen better in right panel of Fig. 3. At high temperatures, that is, closer to the left axis, lnρ is linear on the inverse temperature, i.e. $\ln\rho \sim E_a/k_BT$. The linear fit in the range 100 – 300 K gives $E_a \approx 0.04$ eV. This value correlates reasonably with 0.01 – 0.02 eV observed earlier in $REVO_3$ (RE = La, Y) compounds [30]. Such an energy range is typical for simple activation of charge carriers from the band to the impurity levels near the top of the valence band. At low temperatures, closer to the right axis, the lnρ function is linear on inverse $T^4$, i.e. $\ln\rho \sim (T_0/T)^{1/4}$ indicating for variable range hopping mechanism in three dimensional system [31]. The fit of lnρ in the range 4 – 30 K gives $T_0 = 1.6 \cdot 10^5$ K. The characteristic temperature $T_0$ is connected to the density of states at the Fermi level $N(E_F)$ and charge carrier localization length ξ via ratio $k_BT_0 \approx 18/[N(E_F)\xi^3]$ [32]. By taking the value of ξ about the distance between neighboring V atoms as 5.5 Å the $N(E_F)$ can be estimated as $7.8 \cdot 10^{21}$ eV$^{-1}$cm$^{-3}$. This value is somewhat higher than theoretically calculated density of states in pure $CaVO_3$ $5 \cdot 10^{21}$ eV$^{-1}$cm$^{-3}$[33]. This may indicate for $V^{+3}(d^2)$ state in $CaV_{0.67}W_{0.33}O_3$.

From the data presented, it can be concluded that studied compounds are in the critical region of the metal-insulator quantum phase transition (QPT) [34]. They demonstrate a change in parameters related to the properties of the ground state. Compounds with tungsten content up to 20% are located on the metal side of QPT, and their properties are determined by metal parameters such as carrier concentration and mobility. For compounds containing more than 20-25% of tungsten, the most important parameter is the band gap or activation energy. Among all compounds, the one containing 20% of tungsten has a variable temperature coefficient of resistance, and the absolute value of the resistivity is close to 1 mΩ·cm, which is often used as the Ioffe-Regel limit for the resistivity of metals [35]. Therefore, $CaV_{0.8}W_{0.2}O_3$ compound can be considered as quantum critical point for the metal-insulator transition in this system.

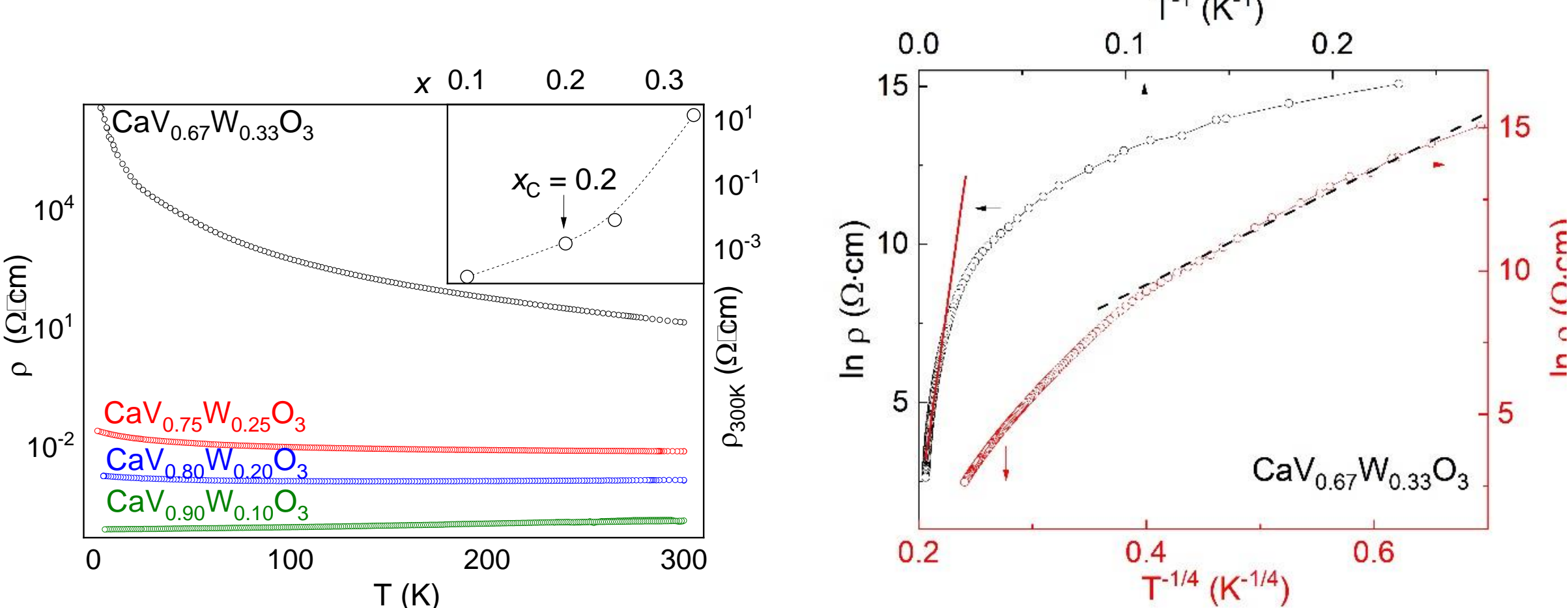

Fig. 3. Left: temperature dependences of resistivity of $CaV_{1-x}W_xO_3$. Inset shows the room temperature resistivity vs. tungsten content x. Right: logarithmic scale of resistivity for $CaV_{0.67}W_{0.33}O_3$ against 1/T (left and top axes) and $1/T^{1/4}$ (right and bottom axes). Solid and dotted lines are the linear fits.

Temperature dependences of magnetic susceptibility χ of $CaV_{1-x}W_xO_3$ are shown in Fig. 4. For low doped samples with $x$ = 0.1 – 0.25 magnetic susceptibility demonstrates temperature independent behavior with a slight increase at low temperatures as shown in left panel of Fig.4.

Temperature independent term can be attributed to the sum of Pauli and Einstein susceptibilities, diamagnetic Pascal's constants and Van – Vleck contribution [36]:

$$\chi_0 = \chi_{Pauli} + \chi_{Einstein} + \chi_{Pascal} + \chi_{VanVleck} \quad (1)$$

Low temperature upturn can be described by Curie Weiss law:

$$\chi_{CW} = \frac{C}{T - \Theta}, \quad (2)$$

with Curie constant $C$ and Weiss temperature Θ. Parameters of the fits of χ(T) with the sum of $\chi_0$ and $\chi_{CW}$ are shown in Table 3.

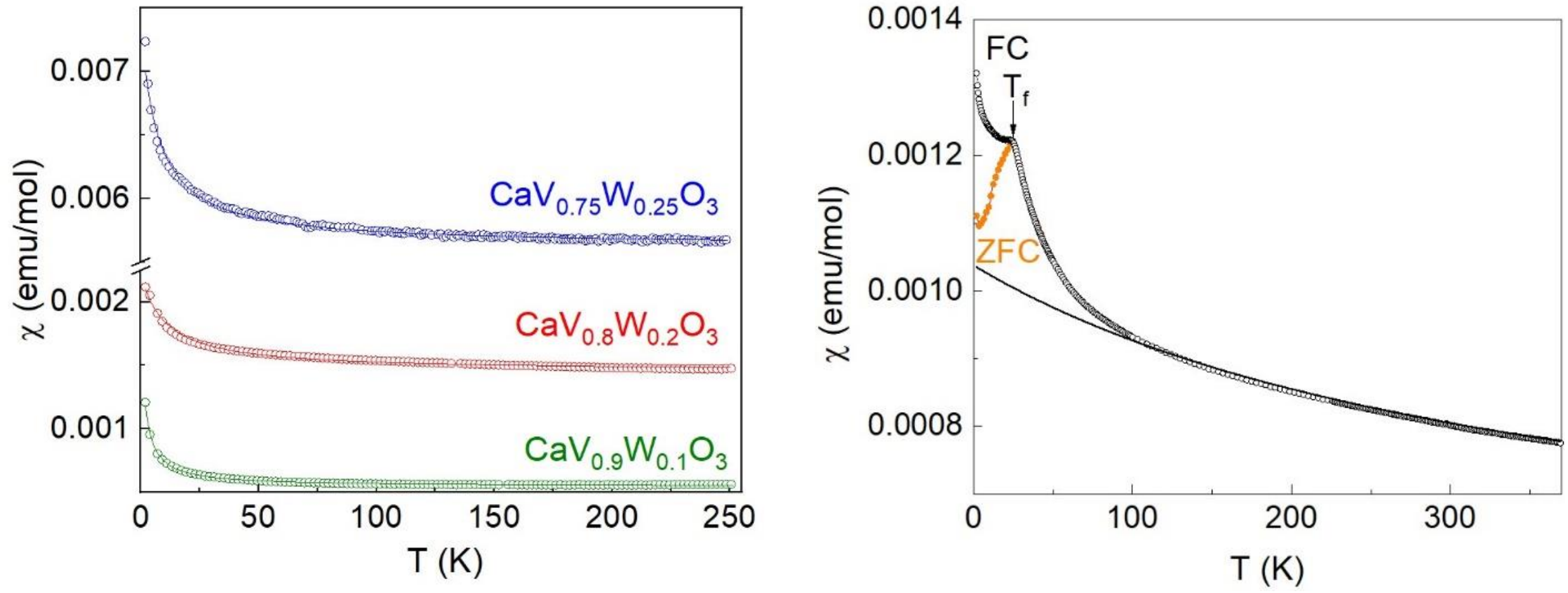

Fig. 4. Left: The temperature dependences of magnetic susceptibility of $CaV_{1-x}W_xO_3$. Right: Temperature dependences of magnetic susceptibility of $CaV_{0.67}W_{0.33}O_3$ measured after cooling in field (FC) and without field (ZFC) at $\mu_0$H=0.1 T. Solid lines are the fits with the sum of $\chi_0$ and $\chi_C$.

Compounds with $x$ = 0.1 – 0.25 possess $\chi_0$ larger than in $CaVO_3$ (~$2.5\cdot10^{-4}$ emu/mol) [31] growing with increasing tungsten content. The $\chi_{Pascal}$ remains almost constant and of the order $-10^{-4}$ emu/mol in $CaV_{1-x}W_xO_3$ row [37]. The Van-Vleck susceptibility of trivalent vanadium, $\chi_{VV}(V^{+3}) \sim 4\cdot10^{-4}$ emu/mol[38], is larger than that of tetravalent ion, $\chi_{VV}(V^{+4}) \sim 6\cdot10^{-5}$ emu/mol [39]. Thus growth of $\chi_0$ up tp $10^{-3}$ emu/mol occurs due to the increase of concentration of $V^{+3}$ ions in $CaV_{1-x}W_xO_3$ with corresponding $\chi_{VV}$ and increase of $\chi_{Pauli} + \chi_{Einstein}$ term, which is proportional to the density of states in Fermi level $N(E_F)$ and effective mass [36].

Table 3. The parameters of the fit of χ(T) of $CaV_{1-x}W_xO_3$ with the sum of $\chi_0$ and $\chi_{CW}$.

| Compound | $\chi_0$ (emu/mol) | $C$ (emu K/mol) | Θ (K) | $\mu_{eff}^{exp2}$, $\mu_B^2$ | $n$ |
|---|---|---|---|---|---|
| $Ca(V^{4+})_{0.7}(V^{3+})_{0.2}W_{0.1}O_3$ | $5.5\cdot10^{-4}$ | 0.002 | -1 | 0.016 | 0.002 |

| $Ca(V^{4+})_{0.4}(V^{3+})_{0.4}W_{0.2}O_3$ | $1.4\cdot10^{-3}$ | 0.008 | -10 | 0.064 | 0.008 |
|---|---|---|---|---|---|
| $Ca(V^{4+})_{0.25}(V^{3+})_{0.5}W_{0.25}O_3$ | $5.6\cdot10^{-3}$ | 0.013 | -7 | 0.104 | 0.014 |
| $Ca(V^{3+})_{0.67}W_{0.33}O_3$ | $5.2\cdot10^{-4}$ | 0.200 | -370 | 1.6 | -* |

* The effective moment cannot be calculated by the spin only value of magnetic moment via equation (4) as described in the text.

Obtained Curie constants allow estimate the square effective moment of localized magnetic moments via equation

$$\mu_{eff}^{exp\,2} = 8C\mu^2{}_B. \qquad (3)$$

These moments are attributed to the impurity $V^{+3}$ ions in pure $CaVO_3$ [36]. Their typical g-factor and spin moment are g=1.91[40] and S= 1. From the square of the effective moment we can estimate the concentration of localized moments *n* via equation:

$$n = \frac{\mu_{eff}^{exp\,2}}{g^2S(S+1)} \qquad . \qquad (4)$$

Low values of *n* indicate that localized moments are attributed to the vanadium defects, which may appear due to deviations in stoichiometry. The values of Θ close to zero indicate for paramagnetic state of low doped samples.

Magnetic susceptibility χ of insulating $CaV_{0.67}W_{0.33}O_3$ grows rapidly with decreasing temperature, demonstrates a cusp at $T_f$ = 24.5 K and a split of dependences measured in field cooled and zero field cooled regimes at $T<T_f$ as is shown in right panel of Fig.4. Such a behavior implies for the formation of a spin-glass state [41]. The parameters of the fit of χ(T) with the sum of $\chi_0$ and $\chi_C$ at 150-370 K are shown in Table 3. Rather low value of $\chi_0$ in comparison with $x$ = 0.1 – 0.25 compounds can be due to Pauli and Einstein terms withdrawal in the absence of mobile charge carriers. Negative Θ signals the presence of antiferromagnetic interactions in the system. Obtained Curie constant allows estimate square effective moment, shown in Table 3, which is less than the theoretical one $\mu_{eff}^{theor\,2} = ng^2S(S+1)\,\mu_B^2 = 4.9\mu_B^2$ μB for $V^{+3}$ with g =1.91, S=1. The reduction of effective moment of a $d^2$ metal is theoretically discussed in distorted octahedral ligand field in Ref. [42]. Experimentally reduction of $V^{+3}$ effective moment is observed in $VI_3$ and explained by subtracting of the orbital moment from the spin only value [43].

The ac-susceptibility of $CaV_{0.67}W_{0.33}O_3$ demonstrates a broad maximum at low temperatures as is shown in left panel of Fig.5. It shifts slightly to higher temperatures for larger frequencies. The measure of the shift $\Delta T_f/(T_f \ln\omega)$ amounts 0.005, which is close to the values typical for interacting cooperative freezing spin-glasses [41,43]. The temperature dependence of relaxation time τ=1/ω is shown in right panel of Fig.5. It obeys the Vogel-Fulcher law: $\tau=\tau_0(\frac{T_f}{T_g}-1)^{zv}$, where $T_f$ is the frequency-dependent freezing temperature determined by the maximum in $\chi_{ac}$, $\tau_0$ – characteristic time scale for the spin dynamics, $T_g$ – glass transition temperature and *zv* – the dynamical exponent, which correlates with the values for the canonical spin glass system CuMn (4.6 at. %) $\tau_0 = 10^{-12}$ s, $zv$ = 7 and $T_g$ = 24.5 K [44].

The temperature dependence of specific heat of $CaV_{0.67}W_{0.33}O_3$, shown in Fig. 6, demonstrates smooth growth without reaching the thermodynamic limit of Dulong - Petite for $m$ = 5 atoms per formula unit and gas constant $R$, $3Rm$ = 125 J/mol K at room temperature. Lattice contribution $C_{lat}$ was approximated by the sum of Einstein [45] and Debye [46] functions, with weights of $a_D$ = 2.6 with $\Theta_D$ = 808 K and $a_E$ = 2.3 with $\Theta_E$ = 286 K. Magnetic contribution to the specific heat $C_m$, shown in the inset to Fig. 6, was obtained by subtraction $C_{lat}$ from $C_p$ due to the insulating state of this compound. It shows a broad maximum at $T_{max}$ = 46 K, which is somewhat higher than $T_f$ [41]. The magnetic entropy obtained by integrating the reduced magnetic heat capacity $C_{magn}/T$ reaches 5.9 J/mol·K, which is close to the theoretical limit $S_m = 4R\ln(2S+1)$ = 6.1 J/mol·K as is shown in the inset to Fig. 6.

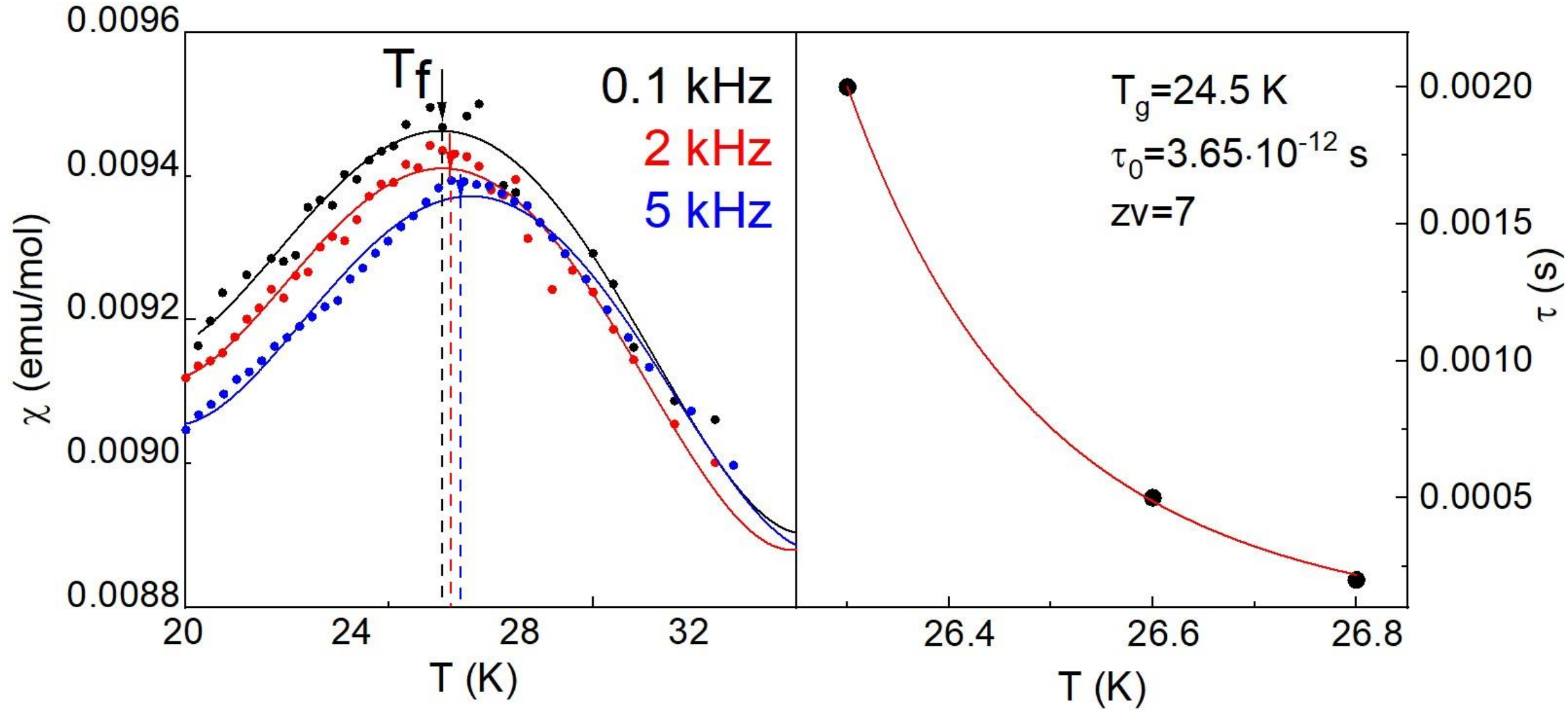

Fig. 5. Left panel: The ac-suspencibility of the $CaV_{0.67}W_{0.33}O_3$ measured at various frequencies. Right panel: The graph of the temperature dependence of relaxation time. Solid line is a fit by the Vogel-Fulcher law.

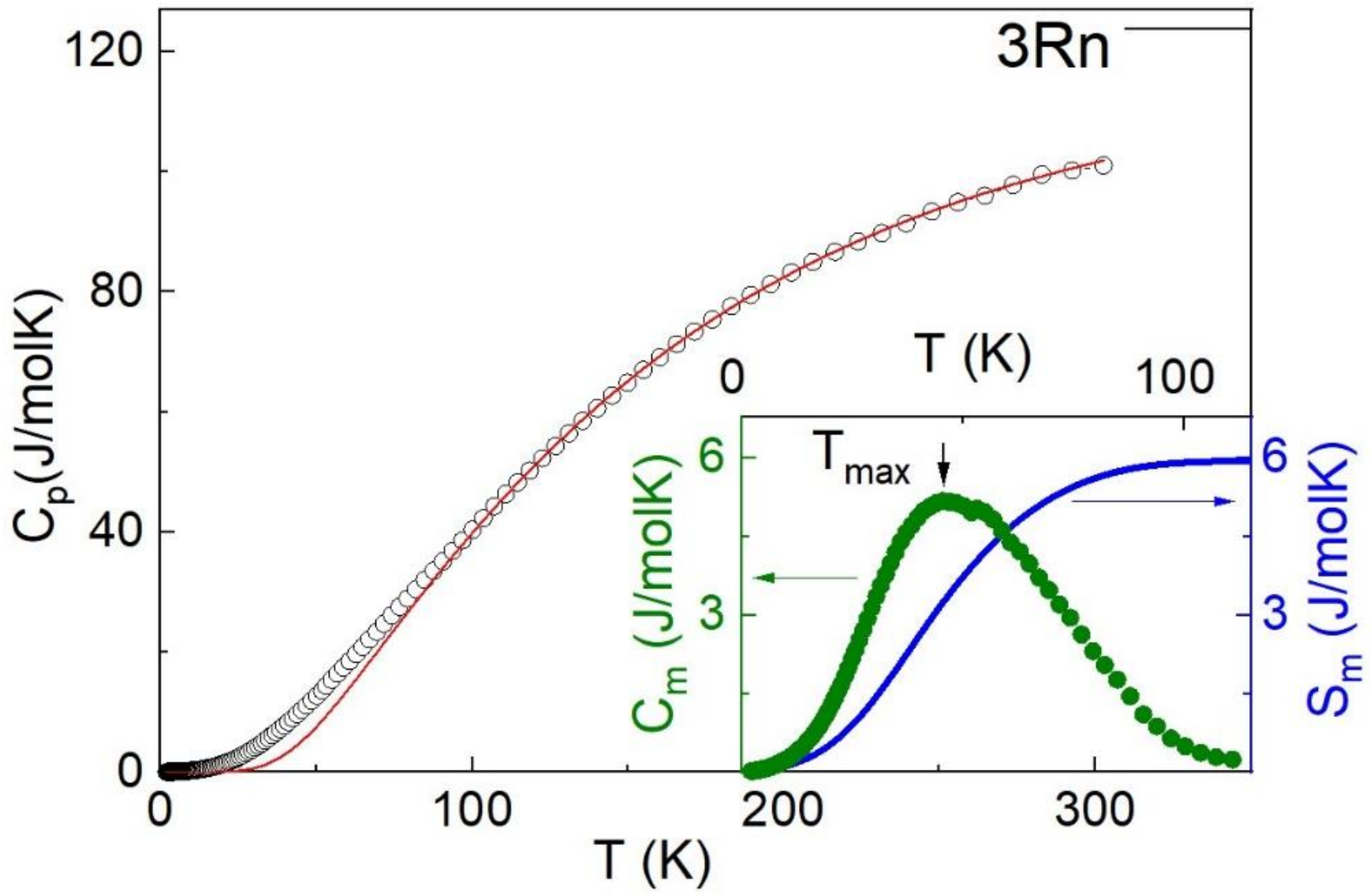

Fig.6. The temperature dependence of specific heat of $CaV_{0,67}W_{0,33}O_3$. Solid line is lattice contribution represented by the sum of Debye and Einstein functions. Horizontal stick is Dulong-Petite limit. The insert represents temperature dependences of magnetic contributions to specific heat $C_m$ and entropy $S_m$.

## Conclusions

A lot of partially ordered and disordered perovskites $AE(B^{+3}_{2/3}W^{+6}_{1/3})O_3$, where B is a triply charged cation, (B = Cr, Mn, Fe, Co, In, Ln), are known to date [47]. The $CaV_{1-x}W_xO_3$ system is an important element of the perovskite puzzle due to the Mott insulator state in the extreme composition with magnetically active $V^{+3}$ ($d^2$) ions, which are rare. All studied samples with $x$ = 0.1 – 0.33 are disordered.

X-Ray diffraction patterns of $CaV_{1-x}W_xO_3$ indicate for the unit cell parameters growth as the tungsten content increases. Distortion of $V/WO_6$ octahedra together with a substantial deviation of the angle between the corner-shared octahedra from 180°, may worsen the metals valence bands overlap with oxygen p – orbitals and induce the localization of electronic states in $CaV_{0.67}W_{0.33}O_3$. Thus steric factor or appearance of "large" $W^{+6}$ and $V^{+3}$ instead of "small" $V^{+4}$ in B position is important for the realization of metal – insulator transition similar to the scenario described in Ref. [16].

The resistivity of $CaV_{0.9}W_{0.1}O_3$ allows to classify it as a metal, compounds with $x$ = 0.2 – 0.25 are between bad metals and semiconductor while x = 0.33 is an insulator. All compounds with low tungsten content $x$ = 0.1 – 0.25 are Pauli paramagnets. Extreme composition $CaV_{0.67}W_{0.33}O_3$ demonstrates magnetic properties, which can be associated with localized magnetic moment on $V^{+3}$ ($d^2$) ions. In accordance with Goodenough – Kanamori – Anderson rules [48], magnetic exchange between $d^2$ metals through the $180^0$ bond via oxygen p – orbitals must be antiferromagnetic. This correlates with antiferromagnetic sign of Weiss temperature Θ. Strongly reduced value of effective moment of $V^{+3}$ is under debate at present and deserves further studies. Due to the structural disorder rather strong antiferromagnetic interactions presented in the system produce magnetically disordered state. At low temperatures $CaV_{0.67}W_{0.33}O_3$ is a canonical spin – glass.

**Acknowledgments**

M.I.V. thanks RSCF 22-43-02020 project for the synthesis and X-Ray analysis. V.O.S. and V.T.M. thank RSCF grant 22-42-08002 for magnetic and thermal measurements. O.E.A. thanks RSCF project 22-72-10034 for transport properties measurements. A.A.A and C.R.G. thanks Ministry of Science and Higher Education of the Russian Federation (agreement no.075-15-2023-448) for XPS measurements

**Author contributions**

**I.V. Morozov:** Investigation, Formal analysis, Writing - original draft **I.K. Shamova** Investigation, Methodology **M.A. Yusifov** Investigation, Methodology **S.Y. Istomin** Conceptualization, Writing - original draft **T.B. Shatalova** Investigation, Methodology **A.I. Boltalin** Investigation, Funding acquisition **A.A. Andreev:** Investigation, Formal analysis R.G. Chumakov: Investigation, Formal analysis **T.M. Vasilchikova:** Investigation, Methodology **A.A. Fedorova** Investigation, Formal analysis **E.A. Ovchenkov** Investigation, Formal analysis **O.S. Volkova:** Funding acquisition, Validation, Writing - review & editing

**Declarations**

**Conflict of interests** The authors declare that they have no conflict of interest.

# Metal – insulator transition in $CaV_{1-x}W_xO_3$ (x = 0.1-0.33) perovskites

I.V. Morozov,[1] I.K. Shamova,[2] M.A. Yusifov,[1,3] S.Y. Istomin,[1,4] T.B. Shatalova,[1] A.I. Boltalin,[1]

A.A. Andreev,[5] R.G. Chumakov,[5] T.M. Vasilchikova,[2,6] A.A. Fedorova,[1] E. A. Ovchenkov,[2,6] O.S. Volkova[2,6]

[1]Department of Chemistry, M. V. Lomonosov Moscow State University, 119991 Moscow, Russian Federation
[2]Functional Quantum Materials Laboratory, National University of Science and Technology "MISiS", Moscow 119049, Russia
[3]Department of Chemistry, Lomonosov Moscow State University, Baku Branch, AZ1146 Baku, Azerbaijan
[4]HSE University, Moscow, Russia
[5]National Research Centre "Kurchatov Institute". 123182, 1 Kurchatov Square, Moscow
[6]Department of Physics, M. V. Lomonosov Moscow State University, 119991 Moscow, Russian Federation

Table S1. Unit cell parameters of polycrystalline $CaV_{1-x}W_xO_3$ compounds obtained in SG *Pnma*.

| Composition | | *a,* Å | *b,* Å | *c,* Å | *V,* Å$^3$ |
|---|---|---|---|---|---|
| Nominal | In accordance with EDX and TG analysis | | | | |
| $CaV_{0.9}W_{0.1}O_3$ | $Ca_{1.02(2)}[V_{0.915(5)}W_{0.085(5)}]_{0.98(2)}O_{3.01}$ | 5.3632(9) | 7.5671(20) | 5.352(4) | 217.21(13) |
| $CaV_{0.8}W_{0.2}O_3$ | $Ca_{1.01(4)}[V_{0.81(2)}W_{0.19(2)}]_{0.99(4)}O_{3.02}$ | 5.4171(17) | 7.6370(11) | 5.3725(11) | 222.26(7) |
| $CaV_{0.75}W_{0.25}O_3$ | $Ca_{1.00(2)}[V_{0.77(2)}W_{0.23(2)}]_{0.99(2)}O_{3.01}$ | 5.442(6) | 7.669(7) | 5.392(6) | 225.1(6) |
| $CaV_{0.67}W_{0.33}O_3$ | $Ca_{1.01(8)}[V_{0.66(6)}W_{0.33(6}]_{0.98(8)}O_{3.01}$ | 5.474(4) | 7.682(5) | 5.405(3) | 227.1(4) |
| $CaV_{0.6}W_{0.4}O_3$* | $Ca_{1.04(4)}[V_{0.64(5)}W_{0.36(5}]_{0.96(4)}O_3$ | 5.478(2) | 7.677(6) | 5.395(3) | 227.3(2) |
| $CaV_{0.5}W_{0.5}O_3$* | - | 5.471(5) | 7.672(6) | 5.398(5) | 226.6(5) |

* $CaV_{1-x}W_xO_3$ samples with $CaWO_4$ impurity.

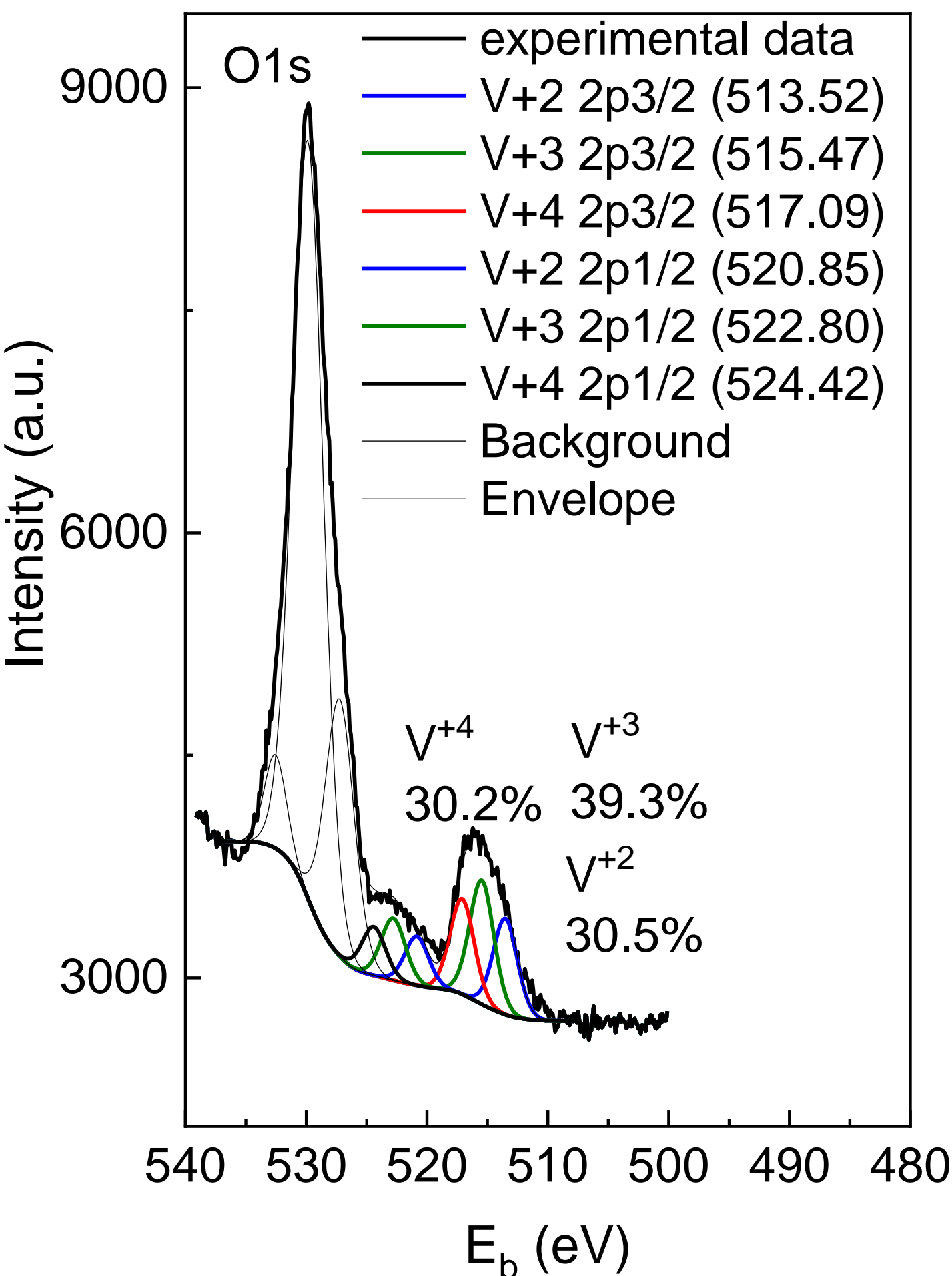

Fig. S1. V 2p (and O 1s) spectrum for $CaV_{0.67}W_{0.33}O_3$.

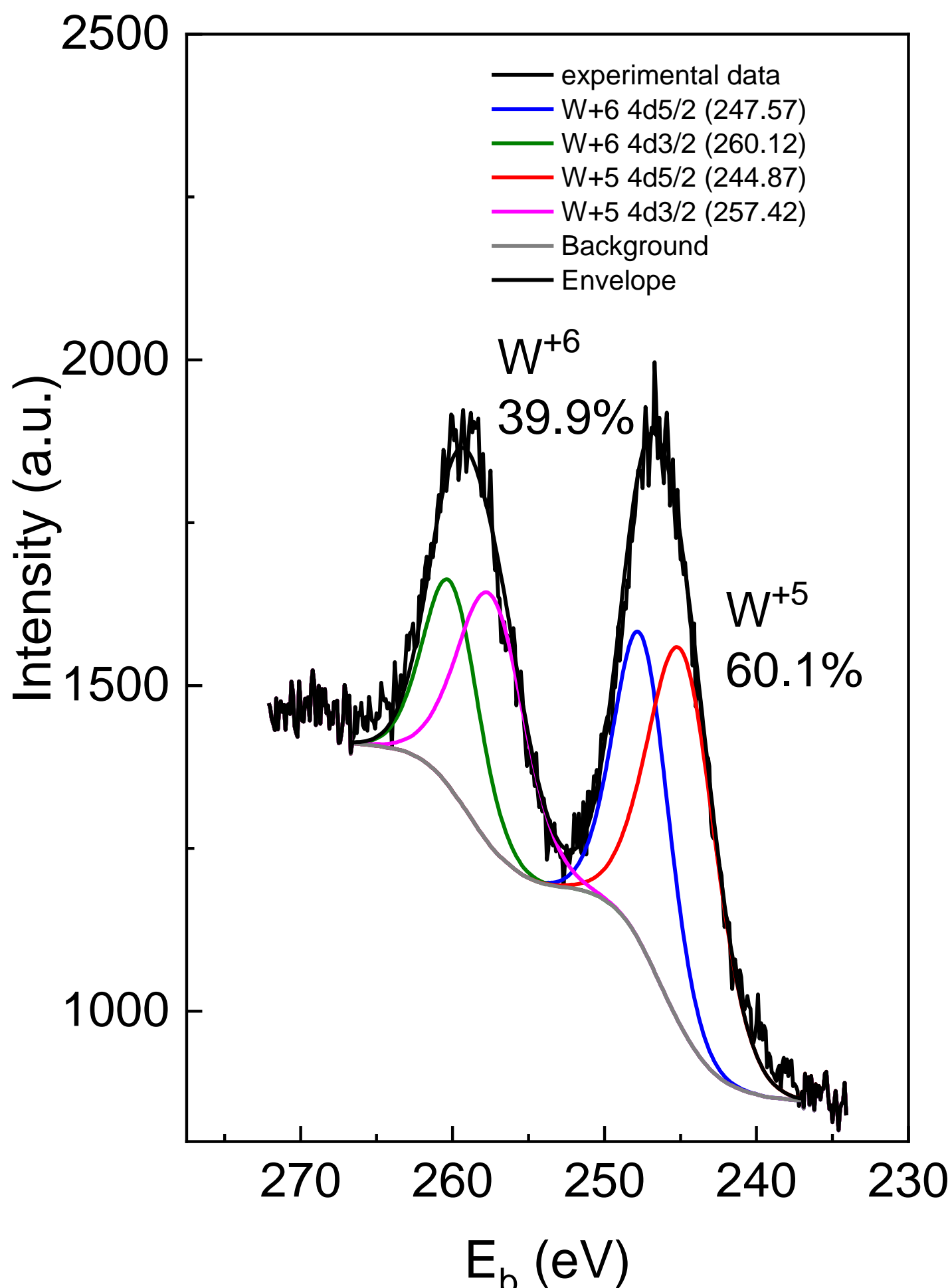

Fig. S2. W 4d spectrum for $CaV_{0.67}W_{0.33}O_3$.

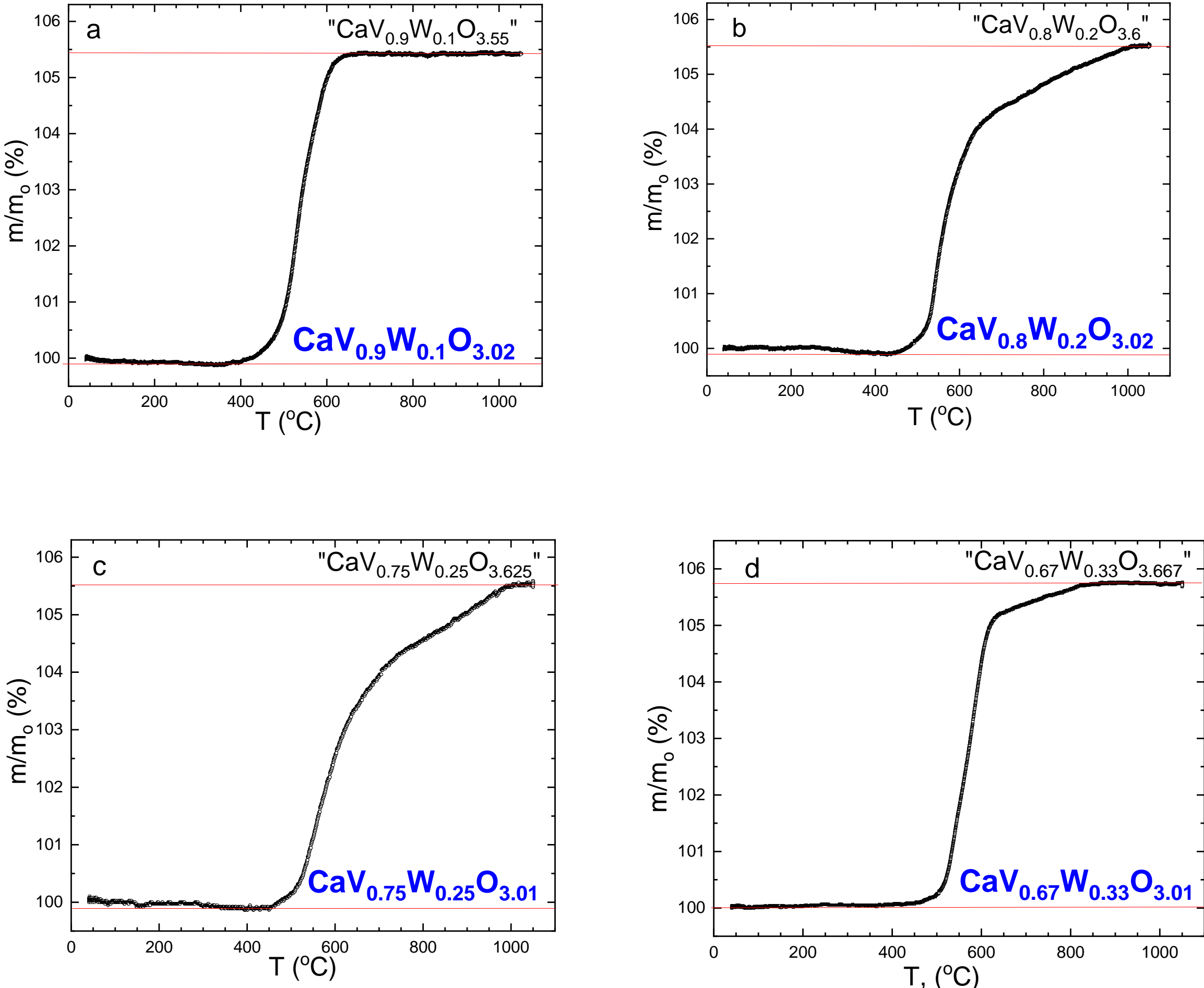

Fig. S3. Thermogravimetric curves for $CaV_{1-x}W_xO_3$ compounds with x = 0.9 (a), 0.8 (b), 0.25 (c) and 0.33 (d). The calculated composition of the oxidized and initial samples are given in quotation marks at the top and in the bottom of figures.

Upon heating up to 1000 °C, the samples oxidize and increase their mass by about 5.1–5.6%. At the final stage of annealing during 1 h at 1050 °C, the mass of the samples does not change. According to XRD data, the final product is a mixture of $Ca_2V_2O_7$ and $CaWO_4$. The molar mass of the initial perovskite was found using the formula:

$$M(CaV_{1-x}W_xO_{3-\delta}) = M(CaV_{1-x}W_xO_{(3.5+0.5x)})\times m_{min}/m_{max},$$

where $CaV_{1-x}W_xO_{(3.5+0.5x)}$ is the molar mass of the oxidized product, $m_{min}$ and $m_{max}$ are the minimum and maximum masses of the sample (in %) on the thermogravimetric curve, respectively. Using the molar mass of the initial perovskite and the degree of substitution x of vanadium with tungsten, we find the amount of oxygen in it. Figure S3 shows that the oxygen content in all the samples is close to the stoichiometric value: $CaV_{1-x}W_xO_3$.

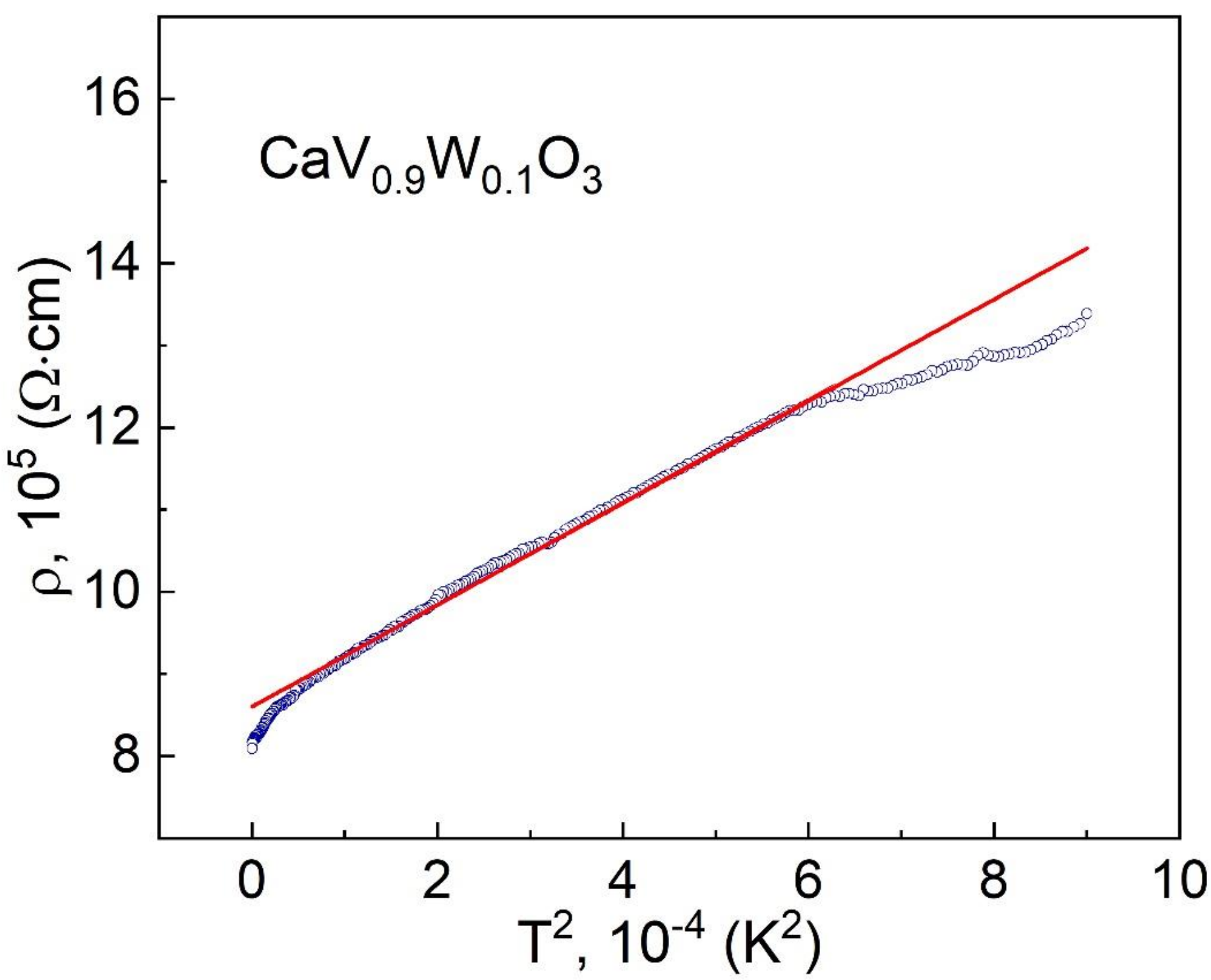

Fig. S4. The dependence of resistivity of $CaV_{0.9}W_{0.1}O_3$ of $T^2$. Solid line is a fit with $\rho = \rho_0 + AT^2$ function.